\RequirePackage{fix-cm}
\documentclass[onecolumn]{svjour3}
\smartqed
\usepackage[utf8]{inputenc}
\usepackage[T1]{fontenc}
\usepackage[hidelinks]{hyperref}
\usepackage{graphicx}
\usepackage{bm}% bold math
\usepackage{mathbbol}
\usepackage{nicefrac}
\usepackage{braket}
\usepackage{calrsfs}
\usepackage{amssymb,amsmath,amsfonts}
\usepackage{cite}
\usepackage{booktabs}
\DeclareMathAlphabet{\mathcal}{OMS}{cmsy}{m}{n}
\SetMathAlphabet{\mathcal}{bold}{OMS}{cmsy}{b}{n}
\newcommand{\mypm}{\mathbin{\smash{%
\raisebox{0.35ex}{%
            $\underset{\raisebox{0.09ex}{$\smash -$}}{\smash+}$%
            }%
        }%
    }%
}

\begin{document}

\title{Control of vertex probability via edge-weight modulation in continuous-time quantum walks}
\titlerunning{Control of vertex probability via edge-weight modulation in CTQWs}

\author{Rafael Vieira  \and Edgard P. M. Amorim}

\institute{R. Vieira   \at
Departamento de F\'isica, Universidade do Estado de Santa Catarina, 89219-710, Joinville, SC, Brazil\\
                           \and
           E. P. M. Amorim \at
Departamento de F\'isica, Universidade do Estado de Santa Catarina, 89219-710, Joinville, SC, Brazil\\
                           \email{edgard.amorim@udesc.br}
}

\date{Received: date / Accepted: date}

\maketitle

\begin{abstract}
Continuous-time quantum walks (CTQWs) provide a versatile framework for exploring quantum transport on graphs. In this work, we investigate how the introduction of edge-weight modulation at a single vertex can suppress its occupation probability. We show that when the edges connected to the root vertex are enhanced by a factor $J$, the probability of detecting the walker at this vertex decays as $1/J^2$, provided the initial state has no components on the vertex itself or its nearest neighbors. We derive the full eigenvalue and eigenvector structure of this system, revealing that the suppression arises from the decoupling of two symmetric line subgraphs and the destructive interference of higher-order contributions. The analysis is extended to tree graphs, where we demonstrate the same scaling behavior and identify the role of local graph geometry in controlling vertex probabilities. These results suggest edge-weight modulation as a mechanism for manipulating transport pathways in CTQWs, with potential applications in quantum information transfer and state engineering, and may serve as a probe of decoherence effects in open quantum systems.
\end{abstract}

%\pacs{03.67.Ac, 03.67.Bg, 05.40.Fb, 05.60.Gg}
%Quantum algorithms, protocols, and simulations & Quantum transport & Entanglement production and manipulation

\maketitle

\section{Introduction} \label{sec:1}

Quantum computation and quantum information are vibrant research fields that have introduced novel models capable of solving certain classes of problems more efficiently than classical algorithms \cite{nielsen2010quantum}. In this context, quantum walks \cite{aharonov1993quantum,farhi1998quantum} have been studied primarily as tools for designing quantum algorithms and protocols \cite{kempe2003quantum,venegas2012quantum}. Quantum correlations in these walks lead to constructive or destructive interference over time. As a result, the quantum walker spreads ballistically, in contrast to the diffusive behavior of a classical particle. This ballistic propagation is the key feature underlying several applications of quantum walks, such as quantum search algorithms \cite{portugal2013quantum}. Beyond these, recent works have identified further applications of quantum walks in teleportation \cite{wang2017generalized,chatterjee2020experimental} and PageRank algorithms \cite{sanchez2012quantum,loke2017comparing,wang2020experimental}. Moreover, quantum walks are also regarded as a promising platform for quantum computation \cite{childs2009universal,lovett2010universal}, feasible in several experimental implementations \cite{wang2013physical,flamini2019photonic}.

Quantum walks are usually presented in two distinct forms: discrete-time \cite{aharonov1993quantum} and continuous-time \cite{farhi1998quantum}. In both cases, the walkers evolve over discrete positions \cite{kempe2003quantum,venegas2012quantum}. The initial state of a discrete-time quantum walk (DTQW) is typically a bipartite state, expressed as a tensor product between a coin state (qubit) and a vertex state. The time evolution is governed by an $\mathrm{SU}(2)$ operator, known as the quantum coin, which generates a new superposition of the qubit. Subsequently, a conditional displacement operation shifts each spin-$1/2$ component in opposite directions toward adjacent positions. In contrast, the initial state of a continuous-time quantum walk (CTQW) consists solely of a vertex state. Driven by a continuous-time Hamiltonian, the walkers evolve according to the Schrödinger equation. Their spreading across the graph is determined by its geometry and the hopping rates between adjacent vertices.

In DTQWs, the engineering of quantum coins has revealed diverse spreading behaviors. On the one hand, uncorrelated random choices of the quantum coin at each time step lead the particle to diffusive spreading while producing maximal entanglement between the coin and position states \cite{vieira2013dynamically,orthey2019weak}. On the other hand, correlated disorder gives rise to a variety of intriguing effects \cite{buarque2019aperiodic,buarque2023discrete,pires2021negative}. When the quantum coins are assigned randomly to the positions, the quantum particle becomes localized \cite{vieira2014entangling}. Indeed, a single distinct quantum coin at the starting position, among Hadamard coins, is sufficient to trap part of the quantum walk state \cite{konno2010localization,wojcik2012trapping,teles2021localization}. Hadamard and $\sigma_x$ gates applied at specific positions and time steps during the walk can induce soliton-like behavior in otherwise delocalized states \cite{zhang2016creating,nitsche2016quantum,orthey2017asymptotic,orthey2019connecting,ghizoni2019trojan}, enabling high-fidelity state transfer \cite{vieira2021quantum,engster2024high}.

The design of graphs has revealed additional applications and effects in CTQWs. For instance, dynamic graphs can provide a pathway for implementing quantum logic gates \cite{herrman2019continuous,wong2019isolated,herrman2022simplifying}. Graph structures with time-dependent weights can localize the state \cite{li2015single} and give rise to Parrondo's effect \cite{ximenes2024parrondo}. Earlier works demonstrated that weighted graphs \cite{zhang2000heavy} can serve as a resource for achieving perfect state transfer \cite{feder2006perfect,canul2009quantum,kendon2011perfect}, universal mixing \cite{carlson2007universal}, and faster quantum search \cite{wong2015faster,wong2016engineering}. More recently, Sett \textit{et al.} investigated the zero transfer effect. This effect consists of the suppression of the probability amplitude at specific vertices in quantum walks on complex-weighted graphs, starting from particular initial states \cite{sett2019Zero} and evolving under the adjacency matrix \cite{wong2016Laplacian}. In subsequent works, Khalique \textit{et al.} and others demonstrated the chiral control of quantum information flux in such walks on complex-weighted graphs, starting from a superposition of multiple vertex states \cite{khalique2021controlled,chaves2023why,bottarelli2023quantum}.

Although the zero transfer effect provides an elegant example of state suppression in quantum walks, it relies on restrictive conditions such as complex edge weights and finely tuned initial states. In this work, we investigate the control of vertex occupation probabilities in CTQWs through local edge-weight modulation. Specifically, we consider tree graphs where the root vertex has edges weighted by a factor $J$ larger than the others. In this configuration, the probability of detecting the walker at the root vertex is not strictly zero but is suppressed as $1/J^2$, provided the initial state has no support on the root or its nearest neighbors. This mechanism demonstrates how simple real-weight modifications can effectively suppress occupation at selected vertices, thereby establishing a practical pathway for controlling quantum transport in more general weighted graphs, with potential applications in quantum state engineering. 

This article is organized as follows. In Sect.~\ref{sec:2}, we present the CTQW model used in this work. In Sects.~\ref{sec:3} and \ref{sec:4}, we calculate the probability at the root vertex for a weighted line and for different types of tree graphs, respectively. In Sect.~\ref{sec:5}, we study these systems under environmental effects. In Sect.~\ref{sec:6}, we present our concluding remarks.

\section{Continuous-time quantum walks}\label{sec:2}

The CTQW model originates from the work of Farhi and Gutmann, who studied a quantum walk on binary trees with the walker starting at the root vertex \cite{farhi1998quantum}. CTQWs are quantum analogs of continuous-time random walks (CTRWs). Whereas CTRWs are governed by a stochastic process, CTQWs describe the dynamics of a particle evolving under the Schrödinger equation over a discrete set of interconnected points, known as a graph. These discrete points are called vertices (or nodes), and the links between them are the edges. Together, the vertices and edges define the structure of a graph.

Formally, a graph $G=(V,E)$ consists of a vertex set $V=\{1,2,\dots,n_v\}$ and an edge set $E=E_u\cup E_w$, where each pair $(i,j)$ connects vertex $i$ to vertex $j$. In this work, the graphs are weighted and undirected, so each pair $(i,j)$ is assigned a weight equal to that of $(j,i)$. This weight $w$ can be physically interpreted as the particle hopping rate from vertex $i$ to vertex $j$. Let us assume the edges $(i,j)\in E_u$ have $w=1$ and $(i,j)\in E_w$ have $w=J$, i.e., $J$ times larger than the weights of edges in $E_u$. Therefore, the adjacency matrix $A$ is an $n_v\times n_v$ matrix defined by
\begin{equation}
A_{ij} = 
\begin{cases}
1, & \quad\text{if } (i,j)\in E_u, \\
J, & \quad\text{if } (i,j)\in E_w, \\
0, & \quad\text{otherwise}.
\end{cases}
\end{equation}
The degree of a vertex $i$ is given by the sum of the elements in the $i$-th row of $A$, and the diagonal degree matrix $D$ is defined as $D_{ii} = \sum_j A_{ij}$.

The quantum state $\ket{\psi(t)}$ of the particle evolves according to the Schr\"{o}dinger equation,
\begin{equation}
i\hbar\frac{\partial}{\partial t}\ket{\psi(t)} = H\ket{\psi(t)},
\end{equation}
where the Hamiltonian $H$ includes a kinetic energy term proportional to the Laplacian of the wave function. If the particle is restricted to discrete positions, such as the vertices of a graph, the Laplace operator is replaced by the graph Laplacian $L=D-A$ \cite{wong2016Laplacian}. In some physical situations, the diagonal degree matrix $D$ can be dropped from $L$. In this case, a common operator choice is the adjacency matrix $A$, which also reflects the geometry of the graph \cite{sett2019Zero,wong2016Laplacian}. Note that both $L$ and $A$ naturally arise in interacting spin models \cite{wong2016Laplacian}. CTQWs driven by the Laplacian and by the adjacency matrix are equivalent on regular graphs, where all vertices have the same degree $\epsilon$. In this case, $D=\epsilon\mathbb{1}$, with $\mathbb{1}$ denoting the identity matrix, and the two time evolutions differ only by an irrelevant global phase. In this work, we focus on CTQWs driven by the adjacency, following previous studies on the zero transfer effect \cite{sett2019Zero,chaves2023why}. Therefore, for an arbitrary initial state $\ket{\psi(0)}$, the time-evolved state is
\begin{equation}
\ket{\psi(\tau)} = e^{-iA\tau}\ket{\psi(0)} 
=\sum_{k=1}^{n_v} e^{-i\lambda_k \tau}\,\braket{\phi_k|\psi(0)}\ket{\phi_k},
\label{Psit}
\end{equation}
where $\lambda_k$ are the eigenvalues and $\ket{\phi_k}$ the corresponding orthonormal eigenvectors of $A$. Here, we introduce the dimensionless time variable $\tau=E_0t/\hbar$, where $E_0$ sets the unit of energy. Throughout this work, the adjacency matrix $A$ is taken to be dimensionless, and the parameter $J$ is a dimensionless factor that simply rescales the edge weights. Since the adjacency governs the dynamics, its spectral properties are fundamental to unravel the evolution of the walk. In particular, eigenvalues and eigenvectors determine how the amplitudes interfere, thereby shaping the probability distribution across the vertices. Consequently, many analytical and numerical studies of DTQWs and CTQWs are based on their spectral decomposition \cite{orthey2017asymptotic,orthey2019connecting,engster2024high}.

\section{CTQWs on a weighted line}\label{sec:3}

\begin{figure}[h]
\centering
\includegraphics[width=0.80\linewidth]{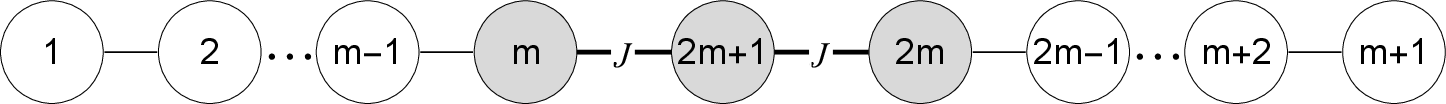}
\caption{Bounded weighted line $L_{2m+1}$: The root vertex $2m+1$ (central) connects to two neighbors with edges weighted by a factor $J$, while all other edges have unit weight. The gray vertices highlight the root vertex and its nearest neighbors.}
\label{Fig1}
\end{figure}

As a starting point, we consider a CTQW on a bounded line with $2m+1$ vertices. The root is the central vertex of the line, but for convenience we relabel it as the last vertex $2m+1$. The edges connected to the root vertex are assigned weights $J$ times larger than those of the remaining edges. The left branch vertices are labeled sequentially from the leaf vertex $1$ up to $m$ (left to right), while the right branch vertices are labeled from $m+1$ up to $2m$ (right to left). Figure~\ref{Fig1} illustrates the bounded weighted line used in our analysis.

In this scenario, the adjacency matrix $A_{2m+1}$ is defined as
\begin{equation}
A_{2m+1} =
\left(
\sum_{i=1}^{m-1}\sum_{j=0}^{1} 
    \ket{mj+i+1}\bra{mj+i}+ J\sum_{k=1}^{2} \ket{2m+1}\bra{mk}
\right) + \mathrm{h.c.}, 
\end{equation}
where $\mathrm{h.c.}$ denotes the Hermitian conjugate of the entire expression inside the parentheses. We first diagonalize $A_{2m+1}$ to determine the set of eigenvalues and their corresponding eigenvectors $\{\lambda_k,\phi_k\}$. For this purpose, we have
\begin{equation}
\Gamma_m = \det\!\left[A_{2m+1}-\lambda\mathbb{1}_{2m+1}\right].
\end{equation}
After a cofactor expansion, we obtain
\begin{equation}
\Gamma_m=-\gamma_m\left(\lambda\gamma_m+2J^2\gamma_{m-1}\right)=0,
\end{equation}
where $\gamma_m=\det[A_{m}-\lambda\mathbb{1}_{m}]$, and $A_{m}$ is the adjacency matrix of an unweighted line graph $L_m$. We thus obtain two conditions: $\gamma_m=0$ and $\lambda\gamma_m+2J^2\gamma_{m-1}=0$. The first condition, $\gamma_m=0$, corresponds to diagonalizing a tridiagonal Toeplitz matrix \cite{Noschese}, which yields the following eigenvalues:
\begin{equation}
\lambda_k=-2\cos\!\left(\frac{k\pi}{m+1}\right), \qquad k=1,2,\ldots,m.
\label{gamma0}
\end{equation}
The recurrence relation $\gamma_m=-\lambda\gamma_{m-1}-\gamma_{m-2}$, when applied to the second condition, implies
\begin{equation}
\left(\lambda^2-2J^2\right)\gamma_{m-1}=\lambda\gamma_{m-2}. 
\label{eq2}
\end{equation}
Furthermore, using the recurrence relation together with the initial conditions $\gamma_0=1$ and $\gamma_1=-\lambda$, we find that
\begin{equation}
\gamma_m = \sum_{k=0}^{\lfloor m/2 \rfloor} 
(-1)^{m+k}\,\frac{(m-k)!}{k!\,(m-2k)!}\,\lambda^{m-2k}, 
\qquad \text{for } m \geq 1.
\label{eq3}
\end{equation}

Although Eq.~\eqref{eq2} admits a formal solution for arbitrary values of $J$, closed analytical expressions that are uniform in both the graph size $m$ and the weight parameter $J$ are not generally accessible. We therefore restrict our analysis to the asymptotic regime $J \gg 1$, where analytical progress is possible and physically relevant. Throughout the manuscript, calculations are carried out within a controlled expansion in $1/J$: terms of order $\mathcal{O}(1/J^3)$ and higher are neglected, while contributions of order $\mathcal{O}(1/J^2)$ are discarded whenever they appear additively with leading $\mathcal{O}(1)$ terms. This approximation is applied consistently in the normalization of eigenvectors and in the derivation of asymptotic expressions for the root occupation probability. At this stage, we propose a solution for the eigenvalues of the form $\lambda \propto J^b$, where $b$ is an arbitrary exponent. We then consider two possible scenarios: $b<1$ and $b \geq 1$. In the first case, where $b<1$, the eigenvalues scale to a value smaller than $J$. In the limit $J \gg 1$, Eq.~\eqref{eq2} reduces to
\begin{eqnarray}
\frac{\gamma_{m-1}}{\gamma_{m-2}}=\frac{\lambda}{\lambda^2-2J^2}\approx-\frac{\lambda}{2J^2}\approx0.
\end{eqnarray}
Therefore, in this regime, where the eigenvalues are of order smaller than $J$, we obtain $\gamma_{m-1}\approx 0$. The $m-1$ eigenvalue solutions arising from this condition can be obtained in the same way as in Eq.~\eqref{gamma0}, yielding the following eigenvalues:
\begin{equation}
\lambda_{m+k} = -2\cos\!\left(\frac{k\pi}{m}\right), 
\qquad k=1,2,\ldots,m-1.
\label{gamma_m-1_0}
\end{equation}
For the second scenario, where $b\geq 1$, we set $\lambda = a J^b$, with $a$ chosen such that both $\lambda \gg 1$ and $J \gg 1$. In this case, Eq.~\eqref{eq2} can be rewritten as
\begin{equation}
a^2 J^{2(b-1)}-2=\frac{a J^{b}}{J^2}\frac{\gamma_{m-2}}{\gamma_{m-1}}.
\label{eq4}
\end{equation}
Similarly, substituting $\lambda=a J^b$ into Eq.~\eqref{eq3} yields
\begin{equation}
\gamma_m=(-1)^m a^m J^{bm}+(-1)^m \sum_{k=1}^{\lfloor m/2 \rfloor} (-1)^k\frac{(m-k)!}{k!(m-2k)!}a^{m-2k} J^{b(m-2k)}.
\label{eq3prime}
\end{equation}
For $J\gg1$ and $b\geq1$, the leading term dominates the summation, as it grows as the highest-order polynomial in $J$. Consequently, we can approximate
\begin{equation}
\frac{\gamma_{m-2}}{\gamma_{m-1}}\approx\frac{(-1)^{m-2}\lambda^{m-2}}{(-1)^{m-1}\lambda^{m-1}}=-\frac{1}{\lambda}=-\frac{1}{a J^{b}}.
\end{equation}
Therefore, substituting this approximation into Eq.~\eqref{eq4}, we obtain
\begin{equation}
2-a^2J^{2(b-1)}\approx \frac{1}{J^2}\xrightarrow[J\gg1]{}0.
\label{eq5}
\end{equation}
Since $b\geq1$, consistency requires $b=1$ and $a=\pm\sqrt{2}$. Hence, the last two eigenvalues are $\lambda_{2m}=-\sqrt{2}J$ and $\lambda_{2m+1}=\sqrt{2}J$. Now that all eigenvalues have been calculated, we proceed to the eigenvectors. The solution for a problem of the type $\gamma_m=0$, corresponding to a 
one-dimensional graph with $m$ vertices, is already known in the literature\cite{Noschese} and is given by
\begin{equation}
\ket{\upsilon_{k}(m)}=\sqrt{\frac{2}{m+1}}
\sum_{j=1}^{m}(-1)^{j}\sin\!\left(\frac{jk\pi}{m+1}\right)\ket{j}, 
\qquad k=1,2,\dots,m.
\end{equation}
When solving the cases of $\lambda_k$ in our model, we must recall that their derivation did not rely on the approximation $1/J^2=0$. Therefore, these eigenvalues are exact rather than approximate. When analyzing the eigenvectors, we observe that their structure resembles two independent line graphs of size $m$, since $\braket{2m+1|\phi_k}\propto \sin(k\pi)/J=0$ for any integer $k$. Thus, the component associated with the root vertex vanishes in these eigenvectors. Consequently, for $\lambda_k$ from Eq.~\eqref{gamma0}, we obtain
\begin{equation}
\begin{aligned}
\ket{\phi_k} 
&= \tfrac{1}{\sqrt{2}}\bigl(\ket{\upsilon_{k}(m)} + \ket{\upsilon_{k}'(m)}\bigr) \\
&= \frac{1}{\sqrt{m+1}}\sum_{j=1}^{m} (-1)^{j} \sin\!\left(\tfrac{jk\pi}{m+1}\right)\bigl(\ket{j}+\ket{j+m}\bigr),
\end{aligned}
\end{equation}
where $\ket{\upsilon_{k}'(m)}$ corresponds to $\ket{\upsilon_{k}(m)}$ with the replacement $\ket{j}\rightarrow\ket{j+m}$. Therefore, these eigenvectors represent symmetric superpositions of two line graphs of length $m$, one on the left branch ($1,\dots,m$) and the other on the right branch ($m+1,\dots,2m$), which evolve 
independently due to the vanishing amplitude at the root vertex. For the eigenvalues $\lambda_{m+k}$ from Eq.~\eqref{gamma_m-1_0}, solving the corresponding eigenvectors yields
\begin{eqnarray}
\ket{\phi_{m+k}} \!&\approx&\! 
\frac{1}{\sqrt{m}}\!
\sum_{j=1}^{m-1}\! (-1)^{j}\sin\!\left(\frac{jk\pi}{m}\right)[\ket{j}\!+\!\ket{m\!+\!j}]
\!+\!\nonumber\\&&\frac{(-1)^{m+k+1}}{\sqrt{m}J}\sin\!\left(\frac{k\pi}{m}\right)\ket{2m\!+\!1}.
\end{eqnarray}

Lastly, for the eigenvalues $\lambda_{2m}=-\sqrt{2}J$ and 
$\lambda_{2m+1}=\sqrt{2}J$, the corresponding eigenvectors are
\begin{equation}
\begin{aligned}
\ket{\phi_{2m}}\!&\approx\! 
\frac{1}{2\sqrt{2}J}\bigl(\ket{m\!-\!1}\!+\!\ket{2m\!-\!1}\bigr)
-\frac{(-1)^{m}}{2}\bigl(\ket{m}\!+\!\ket{2m}\bigr)
+\frac{1}{\sqrt{2}}\ket{2m\!+\!1}, \\[6pt]
\ket{\phi_{2m+1}}\!&\approx\! 
\frac{1}{2\sqrt{2}J}\bigl(\ket{m\!-\!1}\!+\!\ket{2m\!-\!1}\bigr)
+\frac{(-1)^{m}}{2}\bigl(\ket{m}\!+\!\ket{2m}\bigr)
+\frac{1}{\sqrt{2}}\ket{2m\!+\!1}.
\end{aligned}
\label{Phi_2m&2m+1}
\end{equation}
where the normalization assumes $1/J^2=0$. 

Now that the eigenvalues and eigenvectors have been determined, we proceed to investigate the time evolution of the walk, focusing on the probability amplitude at the root vertex to characterize the conditions for the suppression of the occupation probability. We now study the general quantum state of the system,
\begin{equation}
\ket{\psi(\tau)}=\sum_{j=1}^{2m+1}c_j(\tau)\ket{j}.
\end{equation}
For an initial state $\ket{\psi(0)}=\sum_{j=1}^{2m+1}c_j(0)\ket{j}$ and using Eq.~\eqref{Psit}, the probability of finding the walker at the root vertex $r$ is
\begin{equation}
P_r(\tau) = \left|\braket{2m+1|\psi(\tau)}\right|^2
=\left|\sum_{j=1}^{2m+1} e^{-i\lambda_j\tau}\,
\braket{2m+1|\phi_j}\,\braket{\phi_j|\psi(0)}\right|^2,
\label{P_r(tau)}
\end{equation}
where $\sum_{j=1}^{2m+1}\ket{\phi_j}\bra{\phi_j}=\mathbb{1}$ ensures completeness of the eigenbasis. To observe the suppression of the occupation probability at the root vertex $2m+1$, the central amplitude must scale at most as $\mathcal{O}(1/J)$. From the eigenbasis obtained earlier, $\braket{2m+1|\phi_k}=0$ for $k=1,\dots,m$, so these $m$ eigenvectors do not contribute directly to the central amplitude. We therefore focus on $\ket{\phi_{m+k}}$ ($k=1,\dots,m-1$) and on the pair $\ket{\phi_{2m}}$, $\ket{\phi_{2m+1}}$. The latter have central components of magnitude $1/\sqrt{2}$ and would dominate unless their overlaps with the initial state vanish. A sufficient practical condition to suppress $\mathcal{O}(1)$ contributions to the root probability is to choose an initial state with no support on the root vertex and its nearest neighbors, i.e., $|c_m(0)|,~|c_{2m}(0)|,~|c_{2m+1}(0)|\le\varepsilon$, with $\varepsilon\ll 1$. For analytical convenience, we set $\varepsilon=0$ in the calculations. This assumption is justified, since a finite but small $\varepsilon$ introduces only an error of order $\mathcal{O}(\varepsilon/J)$ in the root probability, as demonstrated in Appendix~\ref{sec:apend}. Residual overlaps via vertices $m-1$ and $2m-1$ are $\mathcal{O}(1/J)$ and do not spoil the $1/J^2$ scaling. Accordingly, we take
\begin{equation}
\ket{\psi(0)}=\sum_{\substack{j=1\\ j\neq m}}^{2m-1} c_j\,\ket{j}.
\label{psi0sup}
\end{equation}
For the eigenstates $\ket{\phi_{m+k}}$ ($k=1,\dots,m-1$), first-order large-$J$ analysis yields
\begin{equation}
\braket{2m+1|\phi_{m+k}}\approx\frac{(-1)^{m+k+1}}{\sqrt{m}J}\sin\!\left(\frac{k\pi}{m}\right),
\label{Amp2m+1}
\end{equation}
and
\begin{equation}
\braket{\phi_{m+k}|\psi(0)}=\frac{1}{\sqrt{m}}\sum_{j=1}^{m-1}(-1)^{j}\sin\!\left(\frac{jk\pi}{m}\right)\bigl(c_j+c_{j+m}\bigr). 
\end{equation}
For $\ket{\phi_{2m}}$ and $\ket{\phi_{2m+1}}$, we have
\begin{equation}
\braket{2m+1|\phi_{2m}}=\braket{2m+1|\phi_{2m+1}}\approx\frac{1}{\sqrt{2}},
\label{Amp2m+1_2m}
\end{equation}
and
\begin{equation}
\braket{\phi_{2m}|\psi(0)}=\braket{\phi_{2m+1}|\psi(0)}\approx\frac{1}{2\sqrt{2}J}\bigl(c_{m-1}+c_{2m-1}\bigr).
\end{equation}
Therefore,
\begin{eqnarray}
P_r(\tau)&\approx&\frac{1}{mJ^2}\Bigg|\frac{\sqrt{m}}{2}
\cos\bigl(\sqrt{2}J\tau\bigr)\bigl(c_{m-1}+c_{2m-1}\bigr)+\nonumber\\ && \sum_{k=1}^{m-1} (-1)^{m+k+1}e^{-i\lambda_{m+k}\tau}\,
\sin\!\left(\frac{k\pi}{m}\right)\,
\braket{\phi_{m+k}|\psi(0)}
\Bigg|^2, 
\label{P2m+1}
\end{eqnarray}
with $\lambda_{m+k}=-2\cos\!\left(\frac{k\pi}{m}\right)$. In particular, for the localized initial state $\ket{\psi(0)}=\ket{1}$, we must distinguish between the cases $m=2$ and $m\geq3$. For $m=2$, we have $c_1=1$, while all other coefficients vanish. Thus, we obtain
\begin{equation}
P_r(\tau)\approx\frac{1}{J^2}
\sin^4\left(\frac{J\tau}{\sqrt{2}}\right),\qquad m=2.
\label{Pr1_approx}
\end{equation}
For $m\geq3$, the coefficients $c_{m-1}\!=\!c_{2m-1}=0$. This implies that only the summation in Eq.~\eqref{P2m+1} contributes inside the modulus, and we can rewrite it as

\begin{equation}
P_r(\tau)\approx\frac{1}{m^2J^2}
\left|
\sum_{k=1}^{m-1} (-1)^{k}\,\sin^2\!\left(\frac{k\pi}{m}\right)
\,e^{2i\cos\left(\frac{k\pi}{m}\right)\tau}
\right|^2,\qquad m\geq3.
\label{P2m+1_m>=3}
\end{equation}
Let us consider $m=3$ as an example. In this case, vertices $3$ and $6$ of the two line branches are both adjacent to the root vertex $7$. Using Eq.~\eqref{P2m+1_m>=3}, we obtain
\begin{equation}
P_r(\tau)\approx\frac{\sin^2\!\left(\tau\right)}{4J^2},\qquad m=3.
\label{Pr2_approx}
\end{equation}

In summary, the analysis of the weighted line with $2m+1$ vertices shows that the root vertex exhibits a strongly suppressed occupation whenever the initial state has no support on the root vertex or its nearest neighbors. In this regime, contributions from modes with non-vanishing central amplitudes are reduced, leaving only terms that scale as $1/J^2$. This scaling demonstrates that the probability of detecting the walker at the root vertex can be made arbitrarily small by increasing the relative weight $J$ of the central edges. These results establish the weighted line as a clear example for understanding how local edge-weight modulation can be used to control vertex probabilities in CTQWs, thereby laying the groundwork for the more complex tree graphs considered in the following sections.

\section{CTQWs on weighted tree graphs}\label{sec:4}

Quantum walks have been used to find symmetries in tree graphs \cite{wu2013finding} and to explore paths through various tree structures \cite{reitzner2017finding,koch2018finding}. The trees considered here are defined by having their root vertex as the highest-degree central vertex. From this root emerges a set of connected vertices, called branches. All branches are identical, and the leaf vertices lie at their extremities, each with only one adjacent vertex. Consequently, any state transfer from one leaf vertex to another in a different branch is equivalent and necessarily passes through the root. This feature makes the root vertex a natural candidate for studying mechanisms of probability suppression.

\begin{figure}[h]
\centering
\includegraphics[width=1.0\linewidth]{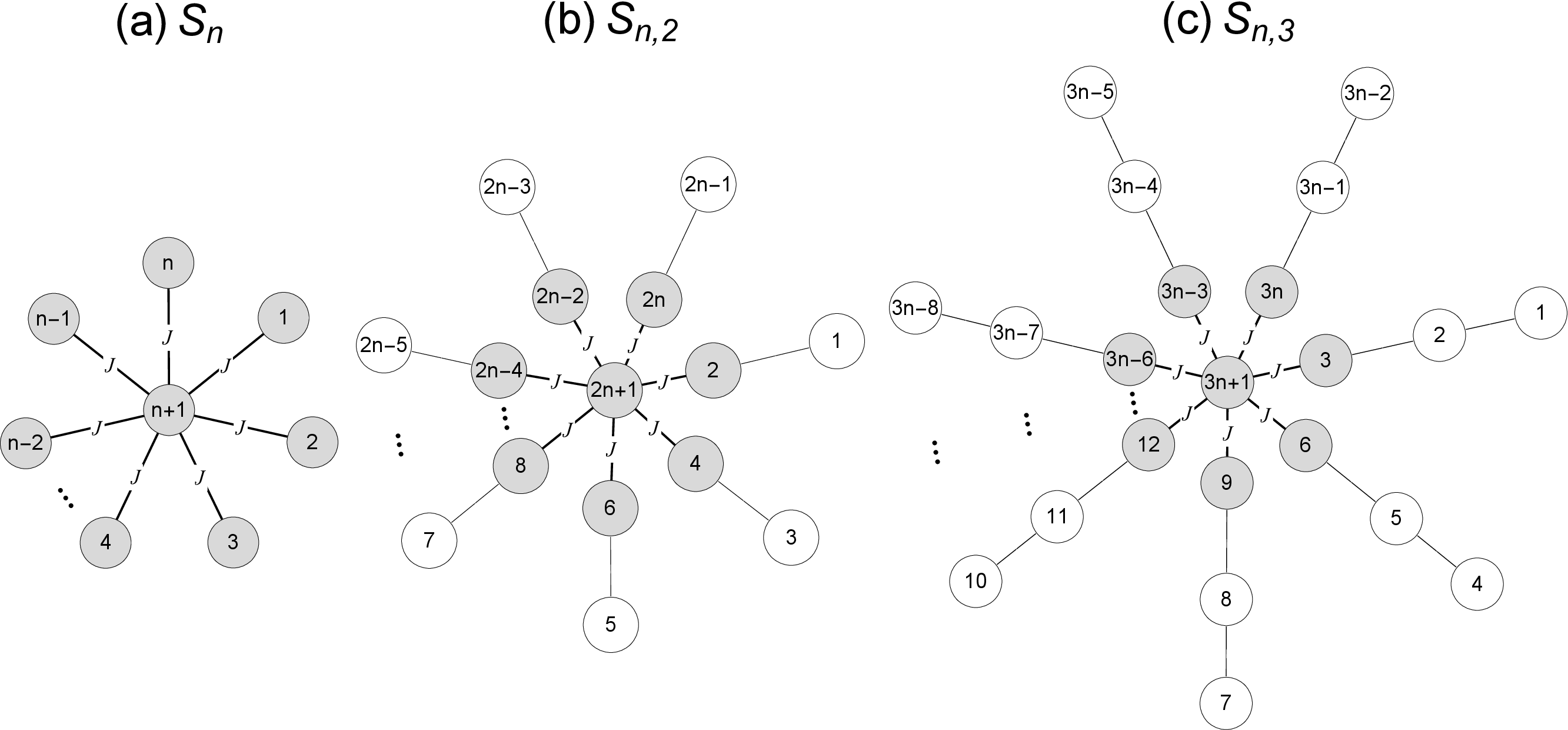}
\caption{Weighted tree graphs: (a) star graph $S_n$, (b) spider graph $S_{n,2}$, and (c) spider graph $S_{n,3}$. 
Gray vertices highlight the root and its $n$ adjacent vertices, whose connecting edges are weighted by $J$.}
\label{Fig2}
\end{figure}

Having established the weighted line as a prototype for local suppression effects, we now turn to weighted tree graphs, which generalize this mechanism to multi-branched structures. In particular, we consider three tree graphs: the star graph $S_n$, and the spider graphs $S_{n,2}$ and $S_{n,3}$, as illustrated in Fig.~\ref{Fig2}. In all cases, the central root vertex connects to $n$ branches, with each branch consisting of $1$, $2$, or $3$ vertices, respectively. The edges incident to the root are weighted by a factor $J$, while all other edges carry unit weight. This construction allows us to explore how the $1/J^2$ suppression of root occupation, observed in the weighted line, extends to more complex topologies. Finally, we establish the equivalence between these graphs and weighted Cayley trees and outline how our main findings extend beyond highly symmetric tree-like structures.

\subsection{Star graphs}

Quantum walks on star graphs have been studied extensively. For instance, Xu showed that quantum walks starting from the root vertex of a star graph exhibit transport behavior equivalent to that of a complete graph of the same size \cite{xu2009exact}. Adding extra edges to star graphs enhances spreading of quantum walks \cite{anishchenko2012enhancing}, and several works have investigated the use of quantum walks to detect structural anomalies in such graphs \cite{feldman2010finding,hillery2012quantum,cottrell2014finding}.

We first examine the root probability $P_r(\tau)$ in a weighted star graph $S_n$ with $n+1$ vertices, where $n$ leaves are connected to a central root vertex by edges of weight $J$, as illustrated in Fig.~\ref{Fig1}(a). The adjacency operator for this case is
\begin{equation}
A_{n+1}=J\sum_{k=1}^{n}|n+1\rangle\langle k|+\mathrm{h.c.},
\end{equation}
and its characteristic polynomial is 
\begin{equation}
\Gamma_n(\lambda)=\det(A_{n+1}-\lambda\mathbb{1}_{n+1})=(-\lambda)^{n-1}(\lambda^{2}-nJ^2)=0. 
\end{equation}
The eigenvalues $\lambda_{1,\dots,n-1}=0$ have associated eigenvectors $\ket{\phi_{1,\dots,n-1}}$ such that $\braket{n+1|\phi_{1,\dots,n-1}}=0$. The remaining eigenvalues are $\lambda_{n}=-\sqrt{n}J$ and $\lambda_{n+1}=\sqrt{n}J$, with corresponding normalized eigenvectors
\begin{align}
\ket{\phi_n}=\frac{1}{\sqrt{2n}}
\begin{bmatrix}
-1 \\ \vdots \\ -1 \\ \sqrt{n} \\
\end{bmatrix}\quad\text{and}\quad
\ket{\phi_{n+1}}=\frac{1}{\sqrt{2n}}
\begin{bmatrix}
1 \\ \vdots \\ 1 \\ \sqrt{n} \\
\end{bmatrix}. 
\end{align}
Since only the eigenstates with nonzero root amplitude contribute, by using Eq.~\eqref{Psit} the root probability evolves as
\begin{equation}
P_r(\tau)=\left|\braket{n+1|\psi(\tau)}\right|^2 =\left|\sum_{k=n}^{n+1} e^{-i\lambda_k \tau}\braket{\phi_k|\psi(0)}\braket{n+1|\phi_k}\right|^2. 
\end{equation}
For an arbitrary initial state given by a superposition of leaves,
\begin{eqnarray}
\ket{\psi(0)}=\sum_{k=1}^{n}a_k\ket{k}, 
\label{psi0lvs}
\end{eqnarray}
where $\sum_k|a_k|^2=1$. By symmetry $P_r(\tau)\propto \left|\sum_{k=1}^{n} a_k\right|^2$. Thus, the root probability vanishes for any initial state in Eq.~\eqref{psi0lvs} whose amplitudes sum to zero, a condition that reflects destructive interference at the root vertex. If the initial state is localized on a single leaf vertex, e.g., $\ket{\psi(0)}=\ket{1}$, the root probability reduces to
\begin{equation}
P_r(\tau)=\frac{\sin^2(J\tau\sqrt{n})}{n}.
\end{equation}
In contrast to the weighted line, where the suppression at the root scales as $1/J^2$, the weighted star graph does not exhibit such suppression: the root probability remains finite and oscillatory, with $J$ only scaling the frequency with which the walker visits the root vertex.

\subsection{Spider graphs}

Adding one or two adjacent vertices to all off-root vertices in $S_n$ generates the spider graphs $S_{n,2}$ and $S_{n,3}$, with $2n+1$ and $3n+1$ vertices, respectively. Both graphs have a root vertex of degree $nJ$. In $S_{n,2}$, the first radial layer consists of vertices of degree $J+1$, while in $S_{n,3}$ an additional second layer of degree-two vertices appears. In both cases, the outermost layer consists of leaf vertices of degree one, as illustrated in Fig.~\ref{Fig2}(b) and (c), respectively. 

In general, the adjacency operator for $S_{n,m}$ is given by
\begin{equation}
A_{mn+1}\!=\!\left(\sum_{i=1}^{m-1}\sum_{j=0}^{n-1}|mj\!+\!i\!+\!1\rangle\langle mj\!+\!i|+J\sum_{k=1}^{n}|mn\!+\!1\rangle\langle mk|\right)\!+\!\mathrm{h.c.},
\label{spider_Adj}
\end{equation}
For $S_{n,2}$ graphs, the characteristic polynomial is $\Gamma_n(\lambda)=\det(A_{2n+1}-\lambda\mathbb{1}_{2n+1})$, which leads to the recursive relation
\begin{equation}
\Gamma_n(\lambda)=(\lambda^2-1)\Gamma_{n-1}+\lambda J^2 (\lambda^2-1)^{n-1}, \qquad n\geq 1.
\label{gammaAQW}
\end{equation}
Using $\Gamma_0=\lambda$, Eq.~\eqref{gammaAQW} simplifies to
\begin{equation}
-\lambda(\lambda^2-1)^{n-1}\left[\lambda^2-(1+nJ^2)\right]=0, \qquad n\geq 1.
\end{equation}
The eigenvalues $\lambda_{1,\dots,n-1}=1$ and $\lambda_{n,\dots,2n-2}=-1$ correspond to eigenvectors with vanishing root component, i.e., $\braket{2n+1|\phi_{1,\dots,2n-2}}=0$. The remaining eigenvalues are $\lambda_{2n-1}=0$, $\lambda_{2n}=\sqrt{1+nJ^2}$, and $\lambda_{2n+1}=-\sqrt{1+nJ^2}$, with normalized eigenvectors
\begin{align}
&\ket{\phi_{2n-1}}=\frac{1}{\sqrt{1+nJ^2}}
\begin{bmatrix} 
-J  \\ 0  \\ \vdots \\ -J       \\ 0         \\ 1 \\
\end{bmatrix},\quad
\ket{\phi_{2n}}=\frac{1}{\sqrt{2n(1+nJ^2)}}
\begin{bmatrix}
1   \\  \sqrt{1+nJ^2} \\ \vdots \\ 1 \\ \sqrt{1+nJ^2}  \\ nJ \\
\end{bmatrix},\quad\text{and} \nonumber \\
&\ket{\phi_{2n+1}}=\frac{1}{\sqrt{2n(1+nJ^2)}}
\begin{bmatrix}
1   \\ -\sqrt{1+nJ^2} \\ \vdots \\ 1 \\ -\sqrt{1+nJ^2} \\ nJ \\
\end{bmatrix}, 
\end{align}
Thus, the root probability evolves as
\begin{equation}
P_r(\tau)=\left|\braket{2n+1|\psi(\tau)}\right|^2=\left|\sum_{j=2n-1}^{2n+1} e^{-i\lambda_{j}\tau}\braket{\phi_j|\psi(0)}\braket{2n+1|\phi_j}\right|^2.
\end{equation}
The initial state of the graph can be chosen arbitrarily, since there is no support on the root vertex or its neighbors. Let us assume $\ket{\psi(0)}=\ket{1}$, we have
\begin{eqnarray}
P_r(\tau)&=&\left|\frac{J}{1+nJ^2}\left(\frac{e^{i\tau\sqrt{1+nJ^2}}+e^{-i\tau\sqrt{1+nJ^2}}}{2}-1\right)\right|^2\nonumber\\
&=&\frac{4J^2}{(1+nJ^2)^2}\sin^4\left(\frac{\tau}{2}\sqrt{1+nJ^2}\right),
\end{eqnarray}
and for $J\gg1$, this expression reduces to
\begin{eqnarray}
P_r(\tau)&\approx&\frac{4}{n^2J^2}\sin^4\left(\frac{\sqrt{n}}{2}J\tau\right),
\label{P2n+1approx}
\end{eqnarray}
then, $J$ scales the frequency in which the walker visiting the root vertex and additionally, this probability also drops with $J^2$, unlike the $S_n$ case. 

For $S_{n,3}$, adding one more vertex to each leaf of $S_{n,2}$ and following the same steps yields the characteristic polynomial
\begin{equation}
-\lambda^{n-1}(\lambda^2-2)^{n-1}\left[\lambda^4-\lambda^2(nJ^2+2)+nJ^2\right]=0.
\end{equation}

The eigenvalues $\lambda_{1,\ldots,n-1}=0$, $\lambda_{n,\ldots,2n-2}=\sqrt{2}$, and $\lambda_{2n-1,\ldots,3n-3}=-\sqrt{2}$ are all associated with eigenvectors that vanish at the root. The remaining four eigenvalues are $\lambda_{3n-2}=\Lambda_+$, $\lambda_{3n-1}=-\Lambda_+$, $\lambda_{3n}=\Lambda_-$, and $\lambda_{3n+1}=-\Lambda_-$, where
\begin{equation}
\sqrt{2}\Lambda_{\pm}=\sqrt{nJ^2+2\pm\sqrt{n^2J^4+4}},
\end{equation}
with their corresponding normalized eigenvectors
\begin{align}
&\ket{\phi_{3n-2}}=
\begin{bmatrix}
\alpha_+ \\ \beta_+ \\ \gamma_+ \\ \vdots \\ \alpha_+ \\ \beta_+ \\ \gamma_+ \\ \delta_+ \\
\end{bmatrix}, 
\quad\ket{\phi_{3n}}=
\begin{bmatrix}
\alpha_- \\ \beta_- \\ \gamma_- \\ \vdots \\ \alpha_- \\ \beta_- \\ \gamma_- \\ \delta_- \\
\end{bmatrix},\quad\text{and}
\quad\ket{\phi_{3n\mp1}}=
\begin{bmatrix}
-\alpha_\pm \\ \beta_\pm \\ -\gamma_\pm \\ \vdots \\ -\alpha_\pm \\ \beta_\pm \\ -\gamma_\pm \\ \delta_\pm \\
\end{bmatrix}, \nonumber
\end{align}
where
\begin{eqnarray}
\alpha_{\mypm}&=&\frac{1}{\sqrt[4]{n^2J^4+4}\sqrt{n\left(nJ^2\mypm\sqrt{n^2J^4+4}\right)}},\nonumber\\
\beta_{\mypm}&=&\frac{J}{\sqrt[4]{n^2J^4+4}\sqrt{nJ^2-2\mypm\sqrt{n^2J^4+4}}},\nonumber\\
\gamma_{\mypm}&=&\frac{\sqrt{nJ^2\mypm\sqrt{n^2J^4+4}}}{2\sqrt{n}\sqrt[4]{n^2J^4+4}},\nonumber\\
\delta_{\mypm}&=&\frac{\sqrt{nJ^2-2\mypm\sqrt{n^2J^4+4}}}{2\sqrt[4]{n^2J^4+4}}.\nonumber\\
\end{eqnarray}
For an initial state localized at a peripheral vertex, e.g. $\ket{\psi(0)}=\ket{1}$, the root amplitude evolves as
\begin{equation}
\braket{3n+1|\psi(\tau)}=-2i\sum_{\sigma=\pm}\alpha_\sigma\delta_\sigma\sin(\Lambda_\sigma \tau),
\end{equation}
and after taking $J\gg1$,
\begin{equation}
\braket{3n+1|\psi(\tau)}\approx -i\left[\frac{\sin(J \tau\sqrt{n})}{n\sqrt{n}J^2}+\frac{\sin \tau}{nJ}\right],
\end{equation}
so that the root probability becomes
\begin{equation}
P_r(\tau)=\left|\braket{3n+1|\psi(\tau)}\right|^2\approx \frac{\sin^2 \tau}{n^2J^2},
\label{P3n+1approx}
\end{equation}
where higher-order terms in $1/J^3$ and $1/J^4$ have been neglected. Notice that Eqs.~\eqref{P2n+1approx} and~\eqref{P3n+1approx} reduce to Eqs.~\eqref{Pr1_approx} and~\eqref{Pr2_approx}, respectively, in the case $n=2$.
In summary, the weighted star graph $S_n$ differs qualitatively from the spider graphs $S_{n,2}$ and $S_{n,3}$. 
While in $S_n$ the parameter $J$ only rescales the oscillation frequency, the spider graphs exhibit suppression of root occupation that scales as $1/J^2$. This contrast demonstrates how the addition of even a single extra vertex per branch fundamentally alters transport properties in CTQWs, transforming the root from a recurrent site of probability flow into one where occupation can be effectively suppressed.

\begin{figure}[h]
\centering
\includegraphics[width=1.0\linewidth]{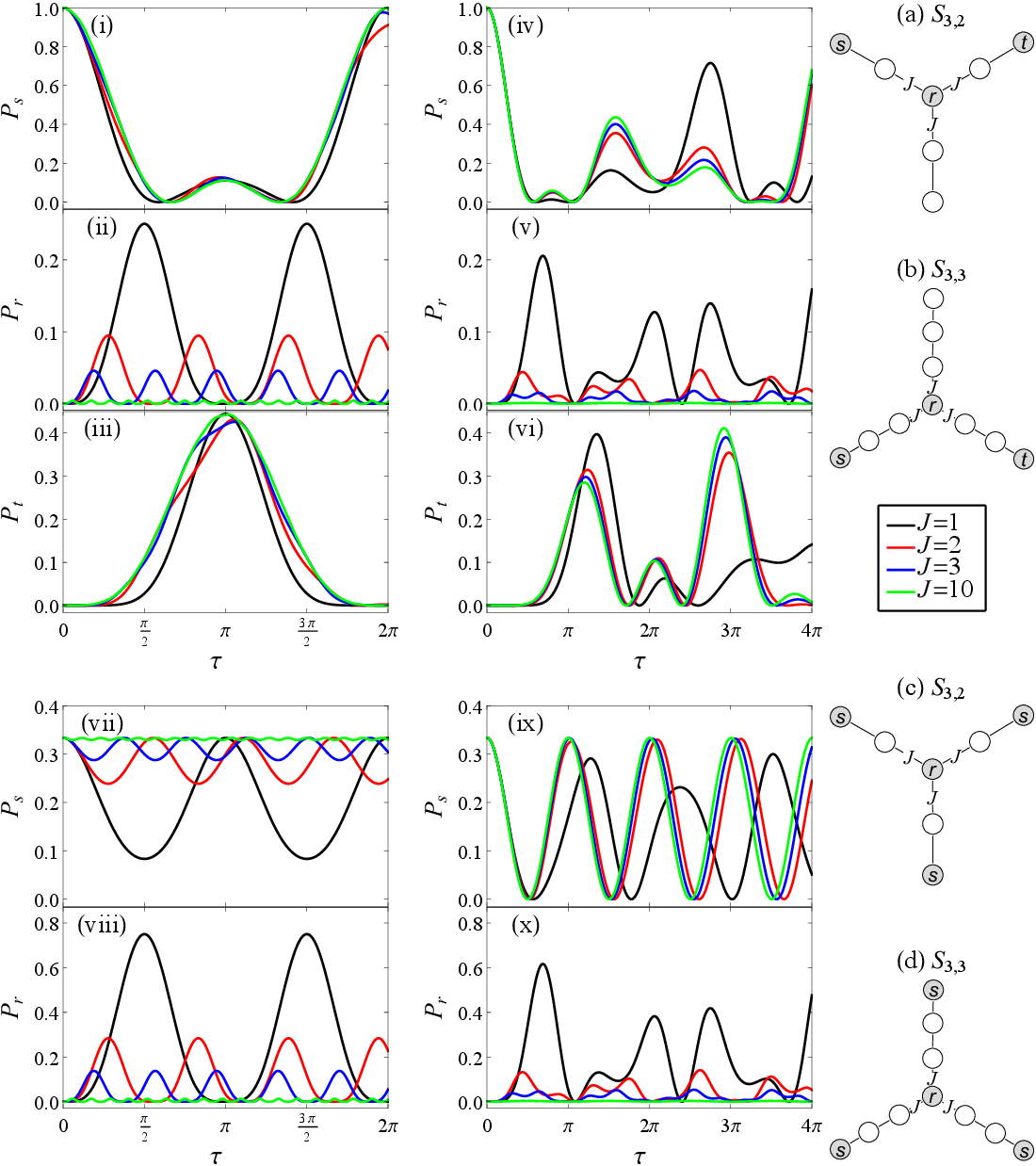}
\caption{Vertex probabilities $P_v(\tau)$ of CTQWs on spider graphs with root-edge weights $J=1$ (black), $2$ (red), $3$ (blue), and $10$ (green). Top: walks starting from a single leaf vertex $s$. Panels (i)–(iii): (a) $S_{3,2}$ showing probabilities at vertex $s$, the root $r$, and a target leaf $t$. Panels (iv)–(vi): analogous results for (b) $S_{3,3}$. Bottom: walks starting from a balanced superposition of the three leaf vertices $s$. Panels (vii)–(viii) for (c) $S_{3,2}$ and panels (ix)–(x) for (d) $S_{3,3}$ showing probabilities at $s$ and $r$. In both cases, the root probability decreases as $1/J^2$.}
\label{Fig3}
\end{figure}

Figure~\ref{Fig3} shows numerical results for vertex probabilities $P_v(\tau)=|\braket{v|\psi(\tau)}|^2$ in CTQWs on spider graphs $S_{3,2}$ and $S_{3,3}$ for different root-edge weights $J$. Panels (i)–(iii) correspond to (a) $S_{3,2}$ and panels (iv)–(vi) to (b) $S_{3,3}$ for walks starting from a single leaf vertex $s$, while panels (vii)–(viii) correspond to (c) $S_{3,2}$ and panels (ix)–(x) to (d) $S_{3,3}$ for walks starting from a balanced superposition of the three leaf vertices. In each case, we track the probability $P_v(\tau)$ at representative vertices $v$: the starting leaf vertex $s$ (or vertices), the root $r$, and a target leaf $t$ on a branch distinct from that containing $s$. Simulations are displayed for increasing values of the central-edge weight $J$.

For both $S_{3,2}$ and $S_{3,3}$, the root probability decreases markedly with increasing $J$, in agreement with the $1/J^2$ scaling derived analytically. In (a) $S_{3,2}$ [panels (i)–(iii)], the walker tends to remain localized near the leaves, with the root probability strongly suppressed as $J$ grows. In (b) $S_{3,3}$ [panels (iv)–(vi)], the extra intermediate layer enriches the dynamics: the state spreads more broadly across the graph, and the return times to the leaves become less sensitive to $J$. This shows how adding even a single vertex per branch qualitatively alters the transport properties while preserving root suppression. A similar pattern appears in (c) $S_{3,2}$ [panels (vii)–(viii)] and (d) $S_{3,3}$ [panels (ix)–(x)], where the initial amplitude is shared among the three leaves, reducing per-leaf probabilities while the root suppression persists.

\begin{figure}[h]
\centering
\includegraphics[width=1.0\linewidth]{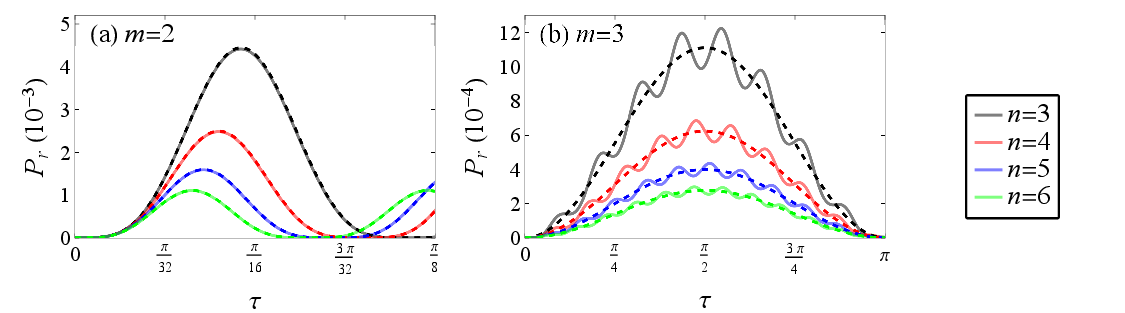}
\caption{Root probability $P_r(\tau)$ at the root $r=mn+1$ for spider graphs with (a) $m=2$, and (b) $m=3$ for branch numbers $n=3$ (black), $4$ (red), $5$ (blue), and $6$ (green). Solid lines are numerical simulations with $J=10$, and dashed lines are the analytical approximations from Eqs.~\eqref{P2n+1approx} and \eqref{P3n+1approx}.}
\label{Fig4}
\end{figure}

Figure~\ref{Fig4} compares the analytical approximations from Eqs.~\eqref{P2n+1approx} and~\eqref{P3n+1approx} with numerical simulations for $J=10$. The plots show the root probability $P_r(\tau)$ for spider graphs with (a) $m=2$ and (b) $m=3$, with branch numbers $n=3,\dots,6$. The dashed curves correspond to asymptotic expressions, which reproduce the exact dynamics with excellent agreement in the large-$J$ regime. These results confirm that the suppression of root occupation in spider graphs is governed by a universal $1/J^2$ law, independent of the branch number $n$.

Finally, we note that zero-transfer conditions at the root of spider graphs have previously been investigated using adjacency operators with complex hopping amplitudes \cite{sett2019Zero}. In contrast, our results demonstrate that a robust suppression of the root occupation can be achieved using purely real edge weights: in the asymptotic regime $J\gg 1$, increasing $J$ alone suffices to make the root probability arbitrarily small. This result holds even when the initial state is an arbitrary superposition of all vertices except the root and its nearest neighbors, without requiring any phase engineering. We thus show that real edge-weight modulation provides an effective mechanism for controlling quantum transport, yielding a root probability that scales as $1/J^2$, with an error-tolerance bound of order $\mathcal{O}(\varepsilon/J)$ (see Appendix~\ref{sec:apend}). In the next section, we discuss the structural equivalence between spider and Cayley trees, thereby extending our analysis.

\subsection{Equivalence to Cayley trees}

\begin{figure}[h]
\centering
\includegraphics[width=0.90\linewidth]{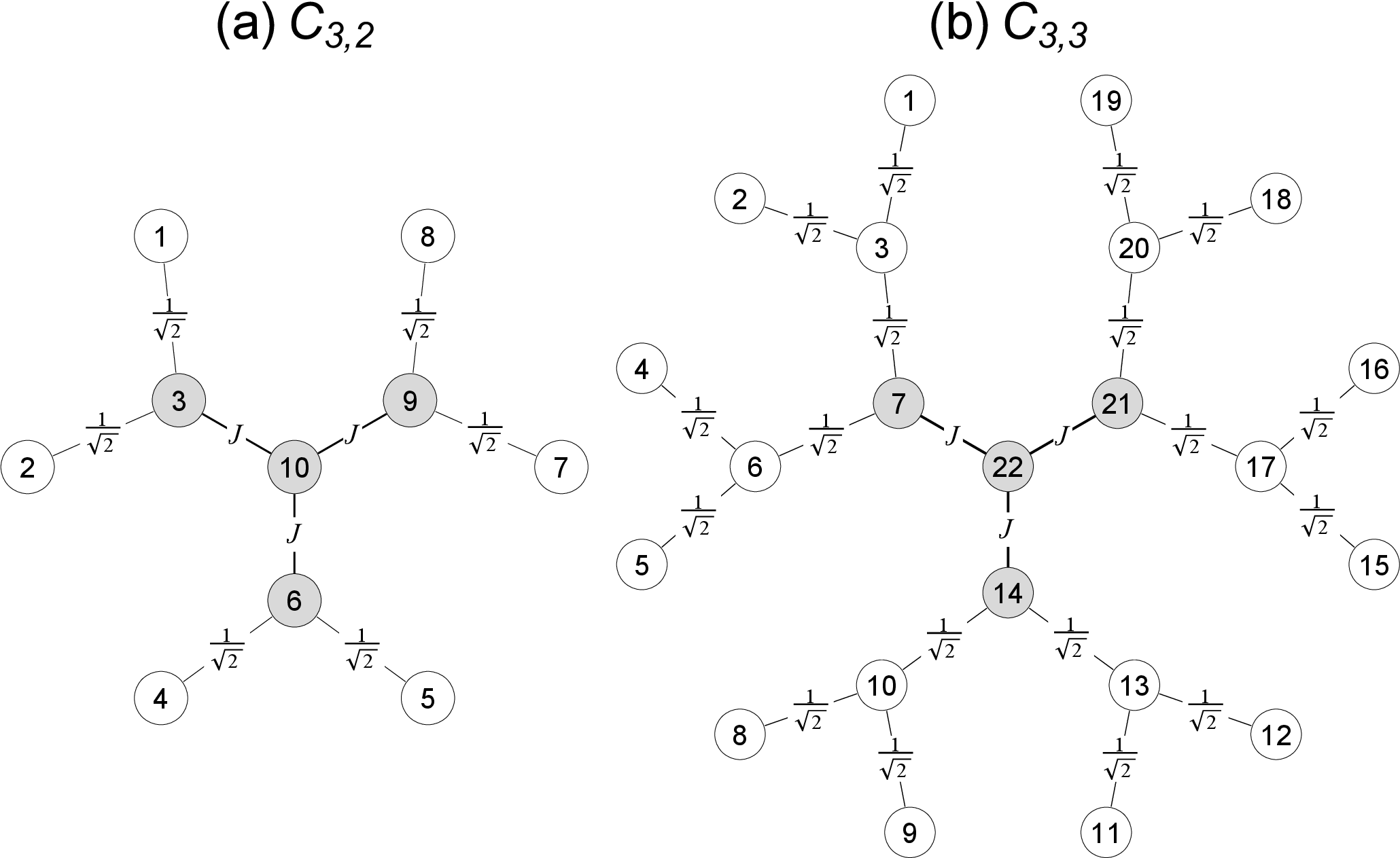}
\caption{Weighted Cayley trees: (a) $C_{3,2}$ and (b) $C_{3,3}$. Their central-vertex probabilities reproduce those of the spider graphs (a) $S_{3,2}$ and (b) $S_{3,3}$ in panels (ii) and (v) of Fig.~\ref{Fig3}, respectively. Edges incident to the root carry weight $J$, while all other edges have weight $1/\sqrt{2}$. Gray vertices indicate the root vertex and its nearest neighbors.}
\label{Fig5}
\end{figure}

Cayley trees $C_{n,m}$ are graphs with $m$ radial layers (levels), where each non-leaf vertex has $n$ adjacent neighbors. We focus on the weighted Cayley trees $C_{3,2}$ and $C_{3,3}$ shown in Fig.~\ref{Fig5}. Their central vertex is connected to three neighbors by edges of weight $J$, analogous to the root of the spider graphs in Fig.~\ref{Fig3}. In these graphs, each first-layer vertex connects to the root (edge of weight $J$) and to its two children by edges of weight $1/\sqrt{2}$. In $C_{3,3}$, the second-layer vertices each connect to three neighbors, also by edges of weight $1/\sqrt{2}$. In both trees, the outermost-layer vertices are leaves with a single edge of weight $1/\sqrt{2}$. In this way, the branches form binary-tree–like structures \cite{farhi1998quantum}. With these weight choices, the initial superposition states $\ket{\psi(0)}=(\ket{1}+\ket{2})/\sqrt{2}$ for $C_{3,2}$ and $\ket{\psi(0)}=(\ket{1}+\ket{2}+\ket{4}+\ket{5})/2$ for $C_{3,3}$ (see Fig.~\ref{Fig5}) reproduce the same root probabilities as the corresponding spider graphs $S_{3,2}$ and $S_{3,3}$ when the walk starts from a single leaf vertex.

Mapping Cayley trees onto spider graphs is feasible through suitable choices of edge weights, as shown above, and this procedure extends naturally to larger graphs. This approach has also been employed in spin chains within the isotropic Heisenberg XX model \cite{vieira2020almost}. Thus, $S_{n,m}$ and $C_{n,m}$ can yield identical root probabilities over time, provided the off-central edges are weighted as $1/\sqrt{n-1}$ and the initial state is a uniform superposition of $(n-1)^{m-1}$ leaf vertices within a single branch. Other choices of weights and initial states can also lead to the same outcome. Different strategies have been proposed in the literature to reduce the dimensionality of regular and non-regular graphs in quantum walks \cite{novo2015systematic,philipp2016continuous}. Therefore, from a qualitative perspective, an arbitrary superposition of leaf vertices in Cayley trees produces the same $J$-dependence, yielding $P_r(\tau)\sim\mathcal{O}(1/J^2)$ for $J\gg1$, as observed in spider graphs.

\subsection{Root suppression beyond symmetric tree-like structures}

The suppression of the root occupation probability relies on two simple local conditions on tree graphs: (i) a root vertex whose incident edges are weighted by a large hopping parameter $J\gg 1$, and (ii) an initial state with negligible amplitude on the root and its nearest neighbors, quantified by a small parameter $\varepsilon\ll 1$ (see Appendix~\ref{sec:apend}). Under these conditions, the strong edge-weight contrast induces an effective decoupling of the root from the rest of the graph, leading to destructive interference that suppresses its occupation probability. This mechanism is therefore local and physical, rather than a consequence of global graph symmetries. Symmetric tree graphs are employed here mainly for analytical convenience, allowing explicit eigensystem derivations summarized in Table~\ref{tab}. 
Nevertheless, the underlying mechanism is expected to persist in more general tree-like structures, provided the local conditions around the root are satisfied.
\begin{table}[htbp]
\centering
\caption{Summary of spectral structure and scaling of the root occupation probability for the graph families studied. Here, $k\in[1,m-1]$ for $L_{2m+1}$ and $\Lambda_{\pm}=\tfrac{1}{\sqrt{2}}\sqrt{nJ^2+2\pm\sqrt{n^2J^4+4}}$ for $S_{n,3}$. Except for the star graph, all graphs exhibit a root probability scaling as $P_r(\tau)\sim 1/J^2$ when the initial state has no support on the root or its nearest neighbors.}
\label{tab:summary}
\begin{tabular}{lll}
\toprule
Weighted graph & Relevant eigenvalues & $P_r(\tau)$ for $\ket{\psi(0)}=\ket{1}$ \\
\midrule
Line $L_{2m+1}$ & $\lambda=\pm\sqrt{2}J$,~$-2\cos\!\left(\tfrac{k\pi}{m}\right)$ & 
$\begin{cases}
\dfrac{1}{J^2}\sin^4\!\left(\tfrac{J\tau}{\sqrt{2}}\right), \quad m=2\\[6pt]
\dfrac{\sin^2(\tau)}{4J^2}, \quad m\geq 3
\end{cases}$ \\ \addlinespace[6pt]
Star graph $S_n$ & $\lambda=0$, $\pm\sqrt{n}\,J$ & $\dfrac{\sin^2(J\tau\sqrt{n})}{n}, \quad n\geq 2$ \\ \addlinespace[6pt]
Spider graph $S_{n,2}$ & $\lambda=\pm 1,\ 0,\ \pm\sqrt{1+nJ^2}$ & $\dfrac{4}{n^2J^2}\sin^4\!\left(\tfrac{\sqrt{n}}{2}J\tau\right), \quad n\geq 2$ \\ \addlinespace[6pt]
Spider graph $S_{n,3}$ & $\lambda=0,\ \pm\sqrt{2},\ \pm\Lambda_{\pm}$ & $\dfrac{\sin^2(\tau)}{n^2J^2}, \quad n\geq 2$ \\
\addlinespace[6pt]
Cayley tree $C_{n,m}$ & Effective reduction to $S_{n,m}$ & Same as $S_{n,m}$ \\
\bottomrule
\end{tabular}
\label{tab}
\end{table}

\begin{figure}[ht]
\centering
\includegraphics[width=0.80\linewidth]{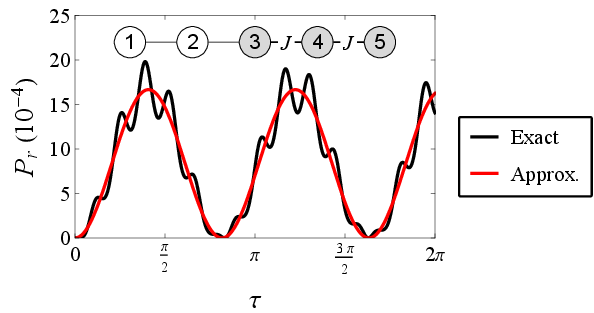}
\caption{Root probability $P_r(\tau)$ for the bounded weighted asymmetric line $L_5$. Exact results (black) are compared with the analytical approximation of Eq.~\eqref{ApproxL5} (red). The root vertex ($r=4$) connects to two neighboring vertices (gray) via edges of weight $J=10$, while all other edges have unit weight.}
\label{Fig6}
\end{figure}

To illustrate the robustness of the suppression mechanism beyond symmetric configurations, we consider an asymmetric weighted line graph, shown in Fig.~\ref{Fig6}. In this bounded weighted line $L_5$, the root vertex (vertex $4$) is displaced from the geometric center and connects to its two neighbors through edges weighted by a factor $J$, while all other edges have unit weight. This configuration is described by the adjacency Hamiltonian
\begin{equation}
A_{5} = \ket{2}\bra{1} + \ket{3}\bra{2} + J\left(\ket{4}\bra{3} + \ket{5}\bra{4}\right) + \mathrm{h.c.},
\end{equation}
which explicitly breaks reflection symmetry. The eigenvalues of $A_5$ are
\begin{equation}
\lambda_{1,2}=\pm \sqrt{1\!+\!J^2\!-\!\sqrt{1\!-\!J^2\!+\!J^4}},~
\lambda_{3,4}=\pm \sqrt{1\!+\!J^2\!+\!\sqrt{1\!-\!J^2\!+\!J^4}},~
\lambda_5=0.
\end{equation}
For $J\gg 1$, we approximate 
\begin{equation}
\sqrt{1-J^2+J^4} = J^2\sqrt{1-\left(\frac{1}{J^2}-\frac{1}{J^4}\right)}\approx J^2-\frac{1}{2},
\end{equation}
which yields $\lambda_{1,2}\approx \pm \sqrt{\frac{3}{2}}$, $\lambda_{3,4}\approx \pm \sqrt{2}\,J$, $\lambda_5=0$. The corresponding eigenvectors are given by
\begin{equation}
\ket{\phi_{1,2}} \approx \frac{1}{\sqrt{12}}
\begin{bmatrix}
2 \\ \pm \sqrt{6} \\ 1 \\ \mp \sqrt{\tfrac{3}{2J^2}} \\ -1
\end{bmatrix},\quad
\ket{\phi_{3,4}} \approx \frac{1}{2}
\begin{bmatrix}
\tfrac{1}{2J^2} \\ \pm \tfrac{1}{\sqrt{2}J} \\ 1 \\ \pm \sqrt{2} \\ 1
\end{bmatrix},~\text{and}\quad
\ket{\phi_5} = \frac{1}{\sqrt{3}}
\begin{bmatrix}
1 \\ 0 \\ -1 \\ 0 \\ 1
\end{bmatrix}.
\end{equation}
The probability of occupying the root vertex $(r=4)$ is given by
\begin{equation}
P_r(\tau) = |\braket{4|\psi(\tau)}|^2 = \left| \sum_{j=1}^{5} e^{-i\lambda_j \tau} \braket{4|\phi_j}\braket{\phi_j|\psi(0)} \right|^2.
\end{equation}
Since $\braket{4|\phi_5}=0$, only the first four eigenstates contribute to the dynamics. Assuming an initial state localized away from the root,
\begin{equation}
\ket{\psi(0)} = c_1\ket{1} + c_2\ket{2}, \qquad |c_1|^2 + |c_2|^2 = 1,
\end{equation}
all contributing terms scale as $1/J$ or $1/J^2$, implying that $P_r(\tau)\sim \mathcal{O}(1/J^2)$. For the specific initial state $\ket{\psi(0)}=\ket{1}$, we obtain
\begin{equation}
P_r(\tau) = \left| \frac{i}{\sqrt{6}J}\sin\!\left(\sqrt{\tfrac{3}{2}}\tau\right)
- \frac{i}{2\sqrt{2}J^2}\sin\!\left(\sqrt{2}J\tau\right) \right|^2.
\end{equation}
Neglecting terms of order $\mathcal{O}(1/J^3)$ and higher, we find
\begin{equation}
P_r(\tau) \approx \frac{1}{6J^2}
\sin^2\!\left(\sqrt{\tfrac{3}{2}}\tau\right).
\label{ApproxL5}
\end{equation}

This simple example demonstrates that the suppression of the root occupation does not rely on graph symmetry. Instead, it emerges from the dynamical decoupling induced by a large local edge-weight contrast, which enforces destructive interference at the root vertex. The mechanism therefore persists in asymmetric and defective tree-like structures, provided the same local conditions on edge weights and initial-state support are satisfied.

\section{Probing decoherence}\label{sec:5}

The time evolution considered so far is unitary, with the quantum walk driven by the adjacency operator $A_{mn+1}$ and thus evolving coherently as a closed system. In realistic settings, however, noise and fluctuations introduce dissipation, scattering, and dephasing \cite{sett2019Zero}, which modify the state and the resulting vertex probabilities. To assess the response at the root vertex in the presence of such effects, we adopt the quantum stochastic walk model as a convenient framework for probing decoherence. This formalism interpolates between coherent quantum evolution and classical stochastic dynamics \cite{whitfield2010quantum,falloon2017qswalk}, allowing one to quantify how incoherent processes progressively suppress interference-driven transport features. By evolving the density operator via the Lindblad master equation
\begin{equation}
\frac{d \hat{\rho}(\tau)}{d\tau}
= -i(1-\omega)\,[A_{mn+1},\hat{\rho}(\tau)]
+ \omega \sum_{k=1}\!\left(\hat{L}_{k}\,\hat{\rho}(\tau)\,\hat{L}_{k}^{\dagger}
-\tfrac{1}{2}\{\hat{L}_{k}^{\dagger}\hat{L}_{k},\hat{\rho}(\tau)\}\right),
\label{lindblad}
\end{equation}
where $0\le\omega\le1$ interpolates between purely coherent ($\omega=0$) and purely incoherent ($\omega=1$) dynamics \cite{whitfield2010quantum}. Environmental scattering between vertices $i$ and $j$ is modeled by Lindblad operators $\hat{L}_{k}$ (note $k$ denotes a directed pair $i\to j$) with $A_{ij}\neq0$ \cite{falloon2017qswalk}, therefore
\begin{equation}
\hat{L}_{i\to j}=\sqrt{|A_{ij}|}\,\ket{i}\!\bra{j},\qquad
A_{ij}=\bra{i}A_{mn+1}\ket{j}. 
\end{equation}
As before, we monitor the probability at the root vertex,
\begin{equation}
P_r(\tau)=\bra{mn+1}\hat{\rho}(\tau)\ket{mn+1}.
\end{equation}

We integrate Eq.~\eqref{lindblad} numerically with a fourth\mbox{-}order Runge--Kutta (RK4) scheme, writing $\dot{\rho}=f(\rho)$ and advancing with step $\Delta\tau$ as
\begin{equation}
\hat{\rho}(\tau_{n+1})
=\hat{\rho}(\tau_{n})
+\frac{\Delta\tau}{6}\,(\kappa_1+2\kappa_2+2\kappa_3+\kappa_4),
\end{equation}
\begin{align}
\kappa_1&=f\!\left(\hat{\rho}(\tau_n)\right),\nonumber\\
\kappa_2&=f\!\left(\hat{\rho}(\tau_n)+\tfrac{\Delta\tau}{2}\,\kappa_1\right),\nonumber\\
\kappa_3&=f\!\left(\hat{\rho}(\tau_n)+\tfrac{\Delta\tau}{2}\,\kappa_2\right),\nonumber\\
\kappa_4&=f\!\left(\hat{\rho}(\tau_n)+\Delta\tau\,\kappa_3\right).
\end{align}

We also compute the cumulative probability at the root,
\begin{equation}
\Omega(\tau)=\int_{0}^{\tau} P_r(\tau')\,d\tau',
\end{equation}
using the initial condition $\hat{\rho}(0)=\ket{1}\!\bra{1}$ and a time step $\Delta\tau=\pi/1000$. In this context, $\Omega(\tau)$ is a convenient quantitative measure of the cumulative population of the root vertex and of the progressive lifting of the suppression induced by decoherence.

\begin{figure}[h]
\centering
\includegraphics[width=1.0\linewidth]{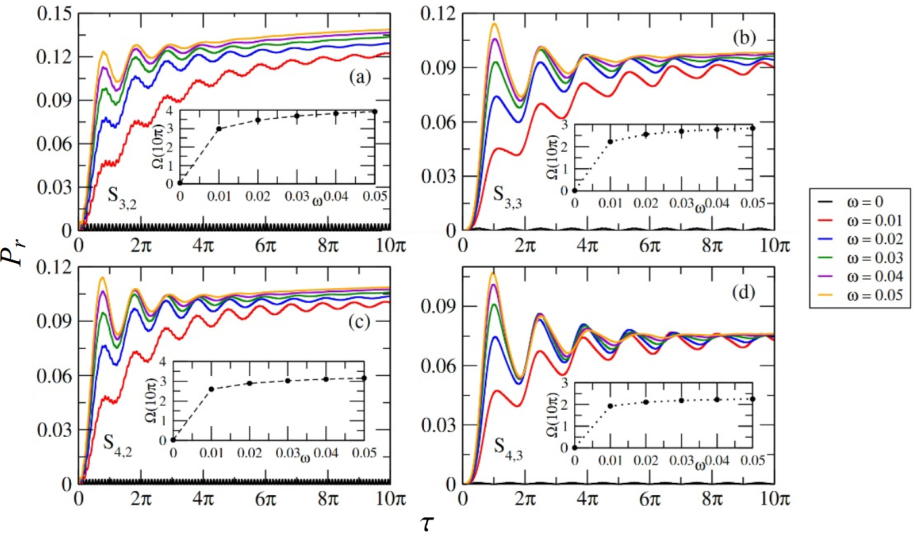}
\caption{Probability at the central vertex $P_r(\tau)$ for spider graphs (a) $S_{3,2}$, (b) $S_{3,3}$, (c) $S_{4,2}$, and (d) $S_{4,3}$ under decoherence, evolved via Eq.~\eqref{lindblad} from $\hat{\rho}(0)=\ket{1}\!\bra{1}$. The decoherence strength is set by $\omega$; insets show the cumulative probability $\Omega(\tau)$ up to $\tau=10\pi$. All simulations use $J=10$ and $\Delta\tau=\pi/1000$. Note that the coherent cases ($\omega=0$, black) remain strongly suppressed compared with $\omega>0$ at long times.}
\label{Fig7}
\end{figure}

Figure~\ref{Fig7} shows $P_r(\tau)$ for several spider graphs as $\omega$ varies from $0$ to $0.05$ in increments of $\Delta\omega=0.01$ (with $J=10$). For any $\omega>0$, the dissipative dynamics drives the system toward the maximally mixed state on the vertex basis, so that $P_r(\tau)\to 1/(mn+1)$ as $\tau\to\infty$ \cite{sett2019Zero}. Thus, different values of $\omega>0$ mainly change the relaxation time to the classical equilibrium, while the coherent case ($\omega=0$) maintains the suppression of the root probability derived earlier. Evaluating $\Omega(\tau)$ for fixed horizons (e.g., $\tau=10\pi$) provides a compact metric of how rapidly decoherence populates the root.

In summary, the suppression of the probability at the root is highly sensitive to decoherence: even weak incoherent scattering ($\omega>0$) washes out the coherent suppression and drives $P_r(\tau)$ toward $1/(mn+1)$. This sensitivity suggests that the root vertex provides a probe of environmental noise in CTQWs on trees, with decoherence progressively decreasing the phase coherence responsible for destructive interference at the root vertex. As a consequence, the suppression mechanism is robust in the coherent regime, but it is gradually lifted as the incoherent contribution increases, driving the root probability toward its classical equilibrium value (uniform distribution). 

\section{Conclusion}\label{sec:6}

We have investigated continuous-time quantum walks (CTQWs) on weighted tree graphs governed by the adjacency operator. We have shown that the probability of occupying the root (central) vertex strongly depends on the hopping strength $J$ of its incident edges and becomes suppressed for large $J$. We first derive the eigensystem of a weighted line, showing that the suppression results from the decoupling of two symmetric subgraphs and destructive interference of higher-order contributions. Extending the analysis to spider and Cayley trees, we have found that the root probability decays as $1/J^2$ for any superposition of vertices taken as the initial state, provided it has no components on the root vertex or its nearest neighbors. This behavior suggests that local edge-weight modulation can robustly suppress root occupation in adjacency-based CTQWs on tree graphs in the regime $J \gg 1$. We have also shown that the root probability is highly sensitive to decoherence, indicating a natural probe of environmental noise in such systems. We emphasize, however, that these conclusions are restricted to adjacency-based dynamics and large edge-weight contrasts, and they may not directly extend to arbitrary graph topologies or other types of quantum-walk dynamics without further investigation.

Since these tree graphs may appear as substructures of larger complex networks \cite{mata2020complex}, they are relevant building blocks for models of quantum transport and communication, and the results presented here may serve as a useful reference for studies of more complex graph structures and experimental studies. In this context, edge-weight modulation could provide a potential mechanism for manipulating transport pathways in CTQWs. Finally, we believe that these results may contribute to the development of efficient algorithms for information transfer, state engineering, and novel probes of decoherence in quantum walks and spin chains \cite{haselgrove2005optimal,bose2007quantum}.

\section*{Acknowledgments}
This work was supported by Conselho Nacional de Desenvolvimento Científico e Tecnológico (CNPq) through grant number 409673/2022-6. R. Vieira thanks Fundação de Amparo à Pesquisa e Inovação do Estado de Santa Catarina 
(FAPESC), Edital 25/2025. E. P. M. Amorim thanks J. Longo for her careful reading of the manuscript.

\appendix

\section{Error-tolerance bound for the root probability} \label{sec:apend}

Recalling Eq.~\eqref{P_r(tau)}, the probability amplitude associated with the root vertex $|2m+1\rangle$ can be written as
\begin{equation}
A_{2m+1}\sim\sum_{j=1}^{2m+1}\braket{2m+1|\phi_j}\braket{\phi_j|\psi(0)}.
\end{equation}
We assume that the initial-state amplitudes on the root vertex and its two nearest neighbors $|c_m(0)|$, $|c_{2m}(0)|$, and $|c_{2m+1}(0)|$ are less than or equal to an arbitrary parameter $\varepsilon$, subject to the normalization condition of the state (hence $\varepsilon\le 1$). 

We decompose the total amplitude as a sum of contributions associated with different eigenvalue sectors of the adjacency operator,
\begin{equation}
A_{2m+1}\sim A^{\lambda_m}_{2m+1}+A^{\lambda_{m+k}}_{2m+1}+A^{\lambda_{2m}}_{2m+1}+A^{\lambda_{2m+1}}_{2m+1}.
\end{equation}
For the eigenvalues
\begin{equation}
\lambda_m=-2 \cos\left(\frac{k\pi}{m+1}\right),
\end{equation}
the corresponding eigenvectors $\ket{\phi_j}$ have no support on the root vertex $\ket{2m+1}$. Therefore, 
\begin{equation}
\braket{2m+1|\phi_j}=0,
\end{equation}
for all $j$ in this sector, which implies 
\begin{equation}
A^{\lambda_m}_{2m+1}=0.
\end{equation}
For the eigenvalues
\begin{equation}
\lambda_{m+k} = -2 \cos\left(\frac{k\pi}{m}\right), 
\end{equation}
Eq.~\eqref{Amp2m+1} shows that the overlap with the root vertex scales as
\begin{equation}
\braket{2m+1|\phi_{m+k}}\sim\frac{1}{J}.
\end{equation}
The factor $\braket{\phi_{m+k}|\psi(0)}$ depends both on the coefficients $c_m(0)$, $c_{2m}(0)$, and $c_{2m+1}(0)$ (each bounded by $\varepsilon$) and on the remaining components of the initial state, which may take arbitrary values. Consequently, the contribution from this sector can be written as
\begin{equation}
A^{\lambda_{m+k}}_{2m+1}=a_1\frac{\varepsilon}{J}+a_2\frac{1}{J},
\end{equation}
where $a_1$ and $a_2$ are complex coefficients independent of $\varepsilon$ and $J$. For the eigenvalues $\lambda_{2m}$ and $\lambda_{2m+1}$, Eq.~\eqref{Amp2m+1_2m} yields
\begin{equation}
\braket{2m+1|\phi_{2m}} =\braket{2m+1|\phi_{2m+1}}\approx\frac{1}{\sqrt{2}}.
\end{equation}
From Eq.~\eqref{Phi_2m&2m+1}, the overlaps $\braket{\phi_{2m}|\psi(0)}$ and $\braket{\phi_{2m+1}|\psi(0)}$ contain one contribution proportional to $\varepsilon$ (originating from $c_m(0)$, $c_{2m}(0)$ and $c_{2m+1}(0)$) and another proportional to $1/J$. Thus,
\begin{equation}
A^{\lambda_{2m}}_{2m+1} + A^{\lambda_{2m+1}}_{2m+1} = a_3\varepsilon+ a_4\frac{1}{J},
\end{equation}
where $a_3$ and $a_4$ are complex coefficients independent of $\varepsilon$ and $J$. Collecting all contributions, and since the root probability is given by $P_r\sim |A_{2m+1}|^2$, we obtain the scaling bound 
\begin{equation}
P_r\sim\left|a_1 \frac{\varepsilon}{J} + (a_2+a_4)\frac{1}{J}+a_3\varepsilon\right|^2.
\end{equation}
Assuming that $J \gg 1$ and $\varepsilon \ll 1/J$, we can neglect $\varepsilon$-dependent corrections of order $\mathcal{O}(\varepsilon^2)$ and $\mathcal{O}(\varepsilon/J^2)$, therefore the root probability scales as
\begin{equation}
P_r \sim \frac{1}{J^2} + \mathcal{O}\left(\frac{\varepsilon}{J}\right).
\end{equation}

Finally, the coefficients $|c_m(0)|$, $|c_{2m}(0)|$, and $|c_{2m+1}(0)|$ need not vanish exactly; it suffices that they are sufficiently small compared with the remaining components of the initial state. For the approximations employed in this work, values $\varepsilon \lesssim 10^{-2}$ are sufficient for these contributions to be safely neglected. This justifies discarding higher-order terms in the $1/J$ expansion (i.e., $\mathcal{O}\left(1/J^3\right)$ and beyond) in the analytical derivations and is consistent with the numerical results shown in Fig.~\ref{Fig3}, which indicate that values as moderate as $J=10$ already make $1/J^2$ corrections small and the suppression mechanism discussed throughout this article manifest clearly.

\end{document}